\documentclass[acmsmall,screen,nonacm]{acmart}

\usepackage{stmaryrd} %

\begin{abstract}

Our work focuses on modeling the security of systems from their component-level designs. Towards this goal, we develop a \emph{categorical} formalism to model attacker actions. Equipping the categorical formalism with algebras produces two interesting results for security modeling. First, using the Yoneda lemma, we can model attacker reconnaissance missions. In this context, the Yoneda lemma shows us that if two system representations, one being complete and the other being the attacker's incomplete view, agree at every possible test, they behave the same. The implication is that attackers can still successfully exploit the system even with incomplete information. Second, we model the potential changes to the system via an exploit. An exploit either manipulates the interactions between system components, such as providing the wrong values to a sensor, or changes the components themselves, such as controlling a global positioning system (GPS). One additional benefit of using category theory is that mathematical operations can be represented as formal diagrams, helpful in applying this analysis in a model-based design setting. We illustrate this modeling framework using an unmanned aerial vehicle (UAV) cyber-physical system model. We demonstrate and model two types of attacks (1) a rewiring attack, which violates data integrity, and (2) a rewriting attack, which violates availability.
\end{abstract}


\title{Yoneda Hacking: The Algebra of Attacker Actions}

\author{Georgios Bakirtzis}
\orcid{0000-0003-4992-0193}
\email{bakirtzis@utexas.edu}
\affiliation{%
  \institution{The University of Texas at Austin}
  \country{USA}
}

\author{Fabrizio Genovese}
\orcid{0000-0001-7792-1375}
\email{fabrizio@statebox.io}
\affiliation{%
  \institution{University of Pisa}
  \country{Italy}
}

\author{Cody H. Fleming}
\orcid{0000-0001-6335-471X}
\email{flemingc@iastate.edu}
\affiliation{%
  \institution{Iowa State University}
  \country{USA}
}

\usepackage{booktabs}
\usepackage{float}
\usepackage{nomencl}
\makenomenclature
\renewcommand{\nompreamble}{Here we summarize some of the symbols we use and their meaning
in category theory as a quick guide for readers
to effectively navigate the following formalism.
}

\usepackage{listings}

\usepackage{etoolbox}
\renewcommand\nomgroup[1]{%
  \item[\bfseries
  \ifstrequal{#1}{C}{Categories}{%
  \ifstrequal{#1}{E}{Equivalences}{%
  \ifstrequal{#1}{O}{Operators}{}}}%
]}

\usepackage{enumerate}

\usepackage{mathtools}

\usepackage{url}

\usepackage{tikz}
\usepackage{tikz-3dplot}

\usepackage[mode=buildnew]{standalone}
\usetikzlibrary{
	math,
	decorations.markings,
	positioning,
	arrows,
	arrows.meta,
	shapes,
	automata,
	petri,
	decorations,
	backgrounds,
	calc,
	fit,
	quotes
	}

\theoremstyle{plain}
\newtheorem{theorem}{Theorem}
\newtheorem{lemma}[theorem]{Lemma}
\newtheorem{corollary}[theorem]{Corollary}

\theoremstyle{definition}
\newtheorem{definition}[theorem]{Definition}
\newtheorem{example}{Example}

\theoremstyle{remark}
\newtheorem*{remark}{Remark}

\newcommand{\CategoryC}{\mathcal{C}}
\newcommand{\CategoryD}{\mathcal{D}}
\newcommand{\CategoryV}{\mathcal{V}}
\newcommand{\CategoryW}{\mathcal{W}}

\newcommand{\Test}{\Theta}

\newcommand{\Reals}{\mathbb{R}}

\usepackage{xcolor}
\definecolor{purp}{RGB}{102, 2, 60}
\definecolor{fgreen}{RGB}{28, 62, 11}

\newcommand{\WD}{\textbf{W}}
\newcommand{\Set}{\textbf{Set}} %
\newcommand{\Cat}{\textbf{Cat}}
\newcommand{\VectR}{\textbf{Vect}_{\Reals}}

\newcommand{\ca}{\mathbf}

\newcommand{\Ob}[1]{\operatorname{obj} \, #1} %
\newcommand{\Hom}[3]{\operatorname{Hom}_{\,#1}\left[#2,#3\right]} %
\newcommand{\Nat}[2]{\operatorname{Nat}\left[#1,#2\right]} %
\newcommand{\Id}[1]{\operatorname{id}_{#1}} %

\newcommand{\inp}{\mathrm{in}}
\newcommand{\out}{\mathrm{out}}

\usetikzlibrary{positioning,decorations.pathreplacing,decorations.pathmorphing,shapes.arrows,decorations.markings,arrows.meta,calc,fit,quotes,cd,math,decorations.markings,arrows,backgrounds,shapes.geometric,shapes}

\newcommand{\SmallBox}[3]
{\begin{tikzpicture}[oriented WD, baseline=-2.5pt, bb Small]
\node[inner sep=.1cm] [bb={1}{1}] (X) {$\scriptstyle #3$};
\draw[label] node[left=.1 of X_in1] (Y) {$#1$}
             node[right=.1 of X_out1] {$#2$};
\end{tikzpicture}}

\tikzcdset{%
   arrow style=tikz,
   diagrams={>={Classical TikZ Rightarrow[angle=63:4pt, line width=.6pt]}},
   arrows={semithick}
}

\tikzset{
  tick/.style={postaction={
    decorate,
    decoration={markings, mark=at position 0.5 with {\draw[-] (0,.4ex) -- (0,-.4ex);}}}
  },
  tickx/.style={
    postaction={ decorate,
      decoration={markings,
        mark=at position 0.5 with {
          \fill circle [radius=.28ex];
        }
      }
    }
  }
}

\tikzset{
   dom/.style={append after command={coordinate[alias=dom#1]}},
   domA/.style={dom=A}, domB/.style={dom=B},
   domC/.style={dom=C}, domD/.style={dom=D},
   domE/.style={dom=E}, domF/.style={dom=F},
   cod/.style={append after command={coordinate[alias=cod#1]}},
   codA/.style={cod=A}, codB/.style={cod=B},
   codC/.style={cod=C}, codD/.style={cod=D},
   codE/.style={cod=E}, codF/.style={cod=F}
}

\tikzset{
   oriented WD/.style={%
      label/.style={
         font=\everymath\expandafter{\the\everymath\scriptstyle},
         inner sep=0pt,
         node distance=2pt and -2pt},
      semithick,
      node distance=1 and 1,
      decoration={markings, mark=at position .5 with {\arrow{stealth};}},
      ar/.style={postaction={decorate}},
      execute at begin picture={\tikzset{
         x=\bbx, y=\bby,
         every fit/.style={inner xsep=\bbx, inner ysep=\bby}}}
      },
   bbx/.store in=\bbx,
   bbx = 1.5cm,
   bby/.store in=\bby,
   bby = 1.75ex,
   bb port sep/.store in=\bbportsep,
   bb port sep=2,
   bb port length/.store in=\bbportlen,
   bb port length=4pt,
   bb min width/.store in=\bbminwidth,
   bb min width=1cm,
   bb rounded corners/.store in=\bbcorners,
   bb rounded corners=2pt,
   bb small/.style={bb port sep=1, bb port length=2.5pt, bbx=.4cm, bb min width=.4cm, bby=.7ex},
   bb Small/.style={bb port sep=1, bb port length=2.5pt, bbx=.5cm, bb min width=.5cm, bby=1ex},
   bb/.code 2 args={%
      \pgfmathsetlengthmacro{\bbheight}{\bbportsep * (max(#1,#2)+1) * \bby}
      \pgfkeysalso{draw,minimum height=\bbheight,minimum width=\bbminwidth,outer sep=0pt,
         rounded corners=\bbcorners,thick,
         prefix after command={\pgfextra{\let\fixname\tikzlastnode}},
         append after command={\pgfextra{\draw
            \ifnum #1=0{} \else foreach \i in {1,...,#1} {
               ($(\fixname.north west)!{\i/(#1+1)}!(\fixname.south west)$) +(-\bbportlen,0) coordinate
               (\fixname_in\i) -- +(\bbportlen,0) coordinate (\fixname_in\i')}\fi %
            \ifnum #2=0{} \else foreach \i in {1,...,#2} {
               ($(\fixname.north east)!{\i/(#2+1)}!(\fixname.south east)$) +(-\bbportlen,0) coordinate
               (\fixname_out\i') -- +(\bbportlen,0) coordinate (\fixname_out\i)}\fi;
         }}}
   },
   bb name/.style={append after command={\pgfextra{\node[anchor=north] at (\fixname.north) {#1};}}}
}

\makeatletter
\def\lst@makecaption{%
  \def\@captype{table}%
  \@makecaption
}
\makeatother

\begin{document}
\maketitle

\section{Introduction}\label{sec:intro}

In the past decade, there has been significant effort in adding formal underpinnings to security modeling~\cite{landwehr:2012,pavlovic:2015}. This research trajectory is evidenced by the NSF/IARPA/NSA workshop on the science of security, which underlines that there are three areas in need of innovation: metrics, formal methods, and experimentation~\cite{evans:2008}. In addition, there is still an increasing need for defining (in)security as a modeling problem~\cite{nicol:2004,avizienis:2004,bakirtzis:dsn:2020}. This paper develops a \emph{formal method} for \emph{modeling} attacker actions at the abstraction level of component-level system models. Specifically, the categorical result of the Yoneda lemma intuitively states that if two system representations agree under any possible test, they behave equivalently. We use this notion to formally show that given two different architectural representations of a system that agree on every test, an attacker can still effectively exploit a system, even with an inaccurate knowledge base of the architecture.

Security engineering has moved from the paradigm of securing a list of assets to modeling in graphs of networked components, which is more congruent with attacker behavior~\cite{lambert_defenders_2015}. These graphs help analyze the system's security posture~\cite{bakirtzis:systems:2019}; however, we can improve them further using the added structure that comes with categorical models of component-level system models, which are by definition compositional; a valuable property for security modeling~\cite{datta:2011}. The basis of the compositional framework lies in functorial semantics, relationships that add meaning to arbitrary syntax based on an already known structure. In the context of security modeling, these functorial semantics that comes with the category-theoretic modeling framework also explicitly relate several essential views for modeling and analyzing cyber-physical systems (CPS), where continuous and discrete behaviors interrelate to produce a total behavior.

The categorical structure comes in the form of decomposition rules between \emph{system behavior} and \emph{system architecture}, which improves upon current practice where system behavior is disjoint from system architecture. Additionally, this approach provides for \emph{early} security modeling, where engineering decisions are most effective~\cite{strafaci_what_2008,bakirtzis:2020}, by operating on models instead of implementations. Early security modeling is possible because categorical semantics reside in a higher level of abstraction than, for example, attack graphs~\cite{sheyner:2002}, which work best after source code is available. Implementing this category-theoretic modeling method allows us to use the Yoneda lemma to show the impact of exploitable vulnerabilities from an attacker's perspective over a system model, which can be incomplete or even partially erroneous compared to the system under attack.

An essential step to building a compositional security modeling framework is reformulating the problem in terms of algebras, where the properties of sets and operations between them are defined precisely. We develop algebras describing behaviors (block diagrams), architectures (graphs), and security operations. We formalize diagrammatic reasoning and unify these differing views of the system (behavior and architecture, and attacker actions). This categorical formalization of diagrammatic reasoning has found success in, for example, manipulating quantum processes~\cite{abramsky:2009,coecke:2010} and databases~\cite{schultz:2016}. We posit a similar innovation would be useful in defining a high-level interpretation of attacker actions diagrammatically. This categorical interpretation formalizes two attacker actions; learning and hijacking. While this framework does not examine particular attacks by itself, attack pattern databases (for example, MITRE CAPEC~\cite{CAPEC}) and attacker modeling frameworks (for example, MITRE ATT\&CK~\cite{strom:2018}) can be used to augment the model with concrete examples of attacker actions.

This paper develops the foundations of security within a \emph{compositional CPS theory} modeling framework~\cite{bakirtzis:2020a}, a flavor of what Lee calls dynamical computational systems theory~\cite{lee:2006}. Formal composition rules could overcome some of the new challenges CPS introduces to cybersecurity~\cite{cardenas:2008,giraldo:2017}, including the intertwined nature of safety and security in this setting. Specifically, our contributions are in the domain of formal methods for security to assist the model-based design of CPS using category theory. These contributions are still bound by well-known problems of the foundations and general \emph{science of security}, such as the lack of a well-defined common language~\cite{herley:2017}.
\begin{itemize}
	\item Describing how fundamental concepts in category theory, such as functors to the category of sets, can be interpreted as testing procedures and results like the Yoneda lemma can be used to infer similarities between systems given the similarities between their test outcomes.
	\item Directly using these insights to model the most common phases of an attack that consist of learning first and on the attack itself afterward, thereby formally modeling from the attacker's perspective.
	\item Extending the approach to CPS by wiring diagrams and their algebras~\cite{bakirtzis:2020}, termed systems-as-algebras, to provide compelling examples of how our mathematical formalization works in practice, which gives rise to formal categorical methods for CPS security.
\end{itemize}

A concrete future result of this line of theoretical work would be the integration of security primitives within systems modeling languages, which will be essential for assuring that CPS models conform to security requirements. We show in a guided process how to achieve this by modeling a rewiring attack and a component attack on an unmanned aerial vehicle (UAV) and how both attacks can be represented diagrammatically as formal methods.

\nomenclature{$\CategoryC$}{The category of categories}
\nomenclature{$\Set$}{The category of sets and functions}
\nomenclature{$\WD$}{The category of labeled boxes and wiring diagrams}
\nomenclature{$\Hom{\CategoryC}{-}{-}$}{Homomorphism within the category $\CategoryC$}
\nomenclature{$\circ$}{Composition (right to left)}
\nomenclature{$\otimes$}{Monoidal product (left to right)}
\nomenclature{$\simeq$}{Isomorphism}
\nomenclature{$f \text{ or } \xrightarrow[]{f}$}{Function}
\nomenclature{$F \text{ or } \xrightarrow[]{F}$}{Functor}
\nomenclature{$F(A) \text{ or } FA$}{Functor application}
\nomenclature{$\Nat{-}{-}$}{Natural transformation}

\nomenclature{$\WD \xrightarrow{F} \Cat$}{Wiring diagram algebra}
\nomenclature{$(X,S)$}{System  with $X \in \WD$ and  $S \in FX$}
\nomenclature{$K_{FX}$}{Knowledge database of systems of type $FX$}
\nomenclature{$FX \xrightarrow{\Theta} \Set$}{Test on $F, X$}
\nomenclature{$\Delta$}{Function duplication}

\section{Problem Formulation}
Having fixed some system, by an \emph{attack}, we mean any procedure intended to change system behavior maliciously. This definition is expansive and ranges from privilege escalation in a computer system to sabotaging a car. We consider an \emph{attacker} any actor who performs any such process to degrade a system's behavior. Traditionally, attacking a system is viewed as an art, and as such, it requires a certain degree of heuristics. Our goal is to formalize these heuristics mathematically, using the attacker's point of view.

It makes sense to divide attacker capabilities into two distinct phases: \emph{learning} and \emph{hijacking}. Learning is where the attacker gathers information about the target system in the hope of finding a weakness. Hijacking is where the attacker exploits a given weakness to change the behavior of some component, which reverberates on the system as a whole. The particulars of how learning and hijacking phases are exercised and how they provide feedback to each other depend on the capabilities and goal of the attacker. In defender terms, this would be the threat model.

To model the learning phase, we need a mathematical description of what probing a system for information means. To model the hijacking phase, we need to mathematically express that the attacker can act on a system to change it.

Crucially, there is an intermediate step between the two phases of how any given weakness is turned into an \emph{exploiting procedure} to change the behavior of some subsystem. The generality of categorical structures models the steps an attacker takes to find system weaknesses and how the exploiting procedures reverberate in the system as a whole. The reason why this abstraction level of modeling security is acceptable is threefold. First, turning a weakness into a viable exploit depends heavily on implementation details, for example, the Spectre and Meltdown exploits~\cite{lipp:2020}. Second, the largest class of attacker actions involves already developed exploits, either internally or through a marketplace, deployed in some sequence to degrade the system's behavior, without necessarily the attacker knowing how they work~\cite{ablon:2014}. Third, suppose a known attack violates a system component. In that case, this intermediate step is unnecessary. At the same time, if it is unknown, the formalism can add increasing levels of detail in the hierarchy to include, for example, source code fragments.

For instance, an attacker may find out by testing (phase 1) that a given laptop uses a particular WiFi card model. The attacker can then purchase -- if it exists -- an exploit for the given card and deploy it. At this point, the system's behavior as a whole will change (phase 2), for example, by giving the attacker the possibility to run any code on the machine. This example is taken straight out of real experience~\cite{Maynor2007}.
Importantly, we claim that an attacker can hijack a system even with an incomplete view of it, as long as the system and the mental model are behaviorally equivalent from the attacker's point of view. Mathematically, this relies on the assumption that the exploit is invariant under 
isomorphism of behavior.

\vspace{1em}
\noindent
\textsc{assumptions} \quad In practice, this means the following: we represent behaviors of systems using the concept of \emph{categorical semantics} (Section~\ref{sec:sas}). This categorical semantics can be more or less granular, depending on how low-level we want our descriptions to be. For instance, consider specifying behavior in terms of automata. We can represent automata as theoretical objects, but we can also view their implementation details. We can describe the system behavior under some formalism and relate it to concrete elements. As a reductionist example, an automaton can be implemented in either a system-on-chip or a field-programmable gate array (FPGA). These would amount to different choices of categorical semantics.

Two automata may be isomorphic in the former setting but not in the latter: this could result in having two different implementations of the same theoretical concept (for example, this is the case in considering the same automaton implemented in two different programming languages). An attack exploiting the automaton design will be isomorphism-invariant in both settings: such an attack exploits the idea that it is possible to start from a state of a given automaton and end up in another state via a legitimate sequence of moves. An attack exploiting the automaton implementation (for example, some weakness of the programming language the automaton is implemented in) will be isomorphism-invariant only in the latter setting. Our categorical semantics cannot ``see'' implementation differences at the theoretical level. This amounts to saying that ``an attack is possible even if the attacker has a differing view of the system, as long as it is behaviorally equivalent to the system itself'' holds only if the categorical semantics developed for application are granular enough to faithfully model the level of generality on which a given exploit acts.

The main mathematical tools used to build the proposed compositional modeling framework for cybersecurity in CPS are: a compositional \emph{system model} in the wiring diagram category; the Yoneda lemma, which represents \emph{tests over the system model}, and more specifically, the Yoneda lemma fully determines a given object by its relationships~\cite{boisseau:2018}; and finally functors and natural transformations to investigate the \emph{effects of an exploit} formally. We have decided to omit from the main text these primitive definitions but have included a formal treatment (Appendix~\ref{sec: preliminaries}) and refer the reader to more comprehensive texts for missing details~\cite{fong:2018,leinster}.

\section{Systems as Algebras}
\label{sec:sas}

We use the categorical structures to develop algebraic machinery such that it is possible to model security violations over a CPS model.
The one assumption we make is that an already existent model of a CPS resides within a category. The motivation for this assumption is twofold. First, through category theory, it is possible to unify behavioral models of control systems with models of candidate concrete implementation of this behavior~\cite{bakirtzis:2020a}. Unification through the categorical formalism traces between algebras and their associated models necessary for the design of CPS~\cite{bakirtzis:2021a}. Second, as it pertains to security assessment, it is helpful to see the effects of an attack at the behavioral level, even though most security assessment techniques apply to implementations -- this relationship builts in the intuition of the security analyst. By having a categorical systems model, we can capture formally and, therefore, precisely the \emph{behavioral} effects of an attack from the particular modeled implementation to the set of control behaviors.

We acknowledge that this assumption contains limitations (Section~\ref{sec: security modelling}), including that systems are often modeled after the fact in a largely informal manner. However, we posit that to move the goal post towards a compositional CPS \emph{theory} as defined in the introduction, we must move towards a more formal treatment of both systems modeling and security analysis. Category theory is but one possible formalism towards this direction. Category theory has the benefit of relating and transforming between different model types. This is an important attribute that can promote and merge existing approaches from the areas of control, systems theory, and security instead of requiring new developments in formalisms for each individual area. 

This view is consistent with the concerns of the industry. Often, unified languages are reductionist in practice, causing increased complexity and lower fidelity of models~\cite{long:2020}. However, the categorical framework we develop here (and others in the area of dynamical systems, for example, by Spivak and Tan~\cite{spivak:2017}) results in the ability to translate between already known and used models. The categorical formalism could be in the backend of a modeling language without the system designer or security analyst interacting with any unfamiliar syntax.

We now use category theory (Appendix~\ref{sec: preliminaries}) to present this compositional theory of CPS. The general systems-as-algebras paradigm was invented by Schultz et al.~\cite{spivak:2016}, and its particular development for CPS was invented by Bakirtzis et al.~\cite{bakirtzis:2020,bakirtzis:2021}. The wiring diagram is the cornerstone category for developing a categorical argument for compositional security analysis for CPS models.

Systems as algebras provide a framework for compositional CPS models that further enable the compositional study of security properties. In applied category theory, diagrams are not mere pictures but instead mathematics in and of themselves~\cite{coecke:2011}. Our choice of categorical structure to represent system models is the category of wiring diagrams and labeled boxes, denoted $\WD$. Therefore, we present the inner workings of the wiring diagram category $\WD$ as explicitly applied to CPS before moving on to address security property modeling.
\begin{figure}[!t]
	\centering
	\includegraphics[width=.5\linewidth]{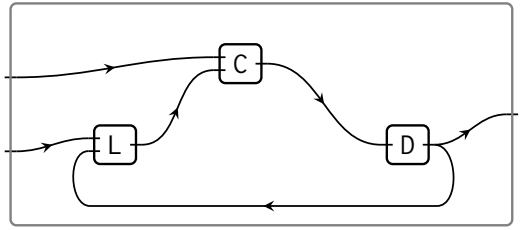}
	\caption{A generic model of a UAV in the wiring diagram category treats the system as black boxes and the connections between them.}\label{fig: UAVAtt}
\end{figure}
\begin{definition}\label{def:w}
	We define the \emph{category of wiring diagrams}, $\WD$, as follows:
	\begin{itemize}
		\item  The objects of the category $\WD$ are pairs of sets $X = (X_\inp, X_\out)$,
		which we depict as labelled boxes.
		
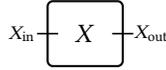
\begin{displaymath}
			\begin{tikzpicture}[oriented WD, bbx=.1cm, bby =.1cm, bb port sep=.15cm,baseline=(X.center)]
				\node [bb={1}{1}] (X) {$X$};
				\draw[label]
				node[left=.1 of X_in1]  {$X_\inp$}
				node[right=.1 of X_out1] {$X_\out$};
			\end{tikzpicture}
		\end{displaymath}
		\item The morphisms of the category $\WD$ are functions.
		Particularly, 
		a morphism $X\to Y$ is a pair of morphisms 
		$( f_\inp\colon X_\out\times Y_\inp\to X_\inp, f_\out\colon X_\out\to Y_\out)$
		in $\Set$ that should be thought of as providing the flow of information in a picture as follows.
		
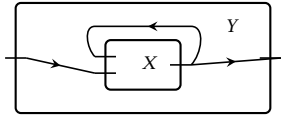
\begin{displaymath}
			\begin{tikzpicture}[oriented WD,baseline=(X.center), bbx=2em, bby=1.2ex, bb port sep=1.2]
				\node[bb={2}{1}] (X) {};
				\node[bb={1}{1}, fit={($(X.north east)+(0.7,1.7)$) ($(X.south west)-(.7,.7)$)}] (Y) {};
				\draw[ar] (Y_in1') to (X_in2);
				\draw[ar] (X_out1) to (Y_out1);
				\draw[ar] let \p1=(X.north west), \p2=(X.north east), \n1={\y1+\bby}, \n2=\bbportlen in
				(X_out1) to[in=0] (\x2+\n2,\n1) -- (\x1-\n2,\n1) to[out=180] (X_in1);
				\draw [label] node at ($(Y.north east)-(.5cm,.3cm)$) {$Y$} node at ($(X.north east)-(.4cm,.3cm)$) {$X$}
				node[above=of Y.north] {};
			\end{tikzpicture}
		\end{displaymath}
	\end{itemize}
 Moreover, the category $\WD$ is monoidal and, therefore, comes equipped with a tensor product $X\otimes Y=(X_\inp\times Y_\inp, X_\out\times Y_\out)$, that formalizes two processes happening in parallel.
	
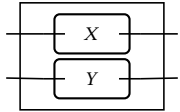
\begin{displaymath}
		\begin{tikzpicture}[oriented WD,baseline=(X2.north), bbx=1.3em, bby=1ex, bb port sep=.06cm]
			\node[bb={1}{1}] (X1) {$\scriptstyle X$};
			\node[bb={1}{1},below =.5 of X1] (X2) {$\scriptstyle Y$};
			\node[fit=(X1)(X2),draw] {};
			\draw (X1_in1) -- (-2.5,0);
			\draw (X1_out1) -- (2.5,0);
			\draw (X2_in1) -- (-2.5,-3.85);
			\draw (X2_out1) -- (2.5,-3.85);
		\end{tikzpicture}
	\end{displaymath}
\end{definition}
The wiring diagram category, $\WD$, gives a formal composition rule for connecting labeled boxes and wires. To add some meaning to those boxes, we need to develop an algebra that assigns some form of behavior, for example, in the form of automata or state-space models.
The semantics of CPS will predominantly exist in the category of sets and functions $\ca{Set}$ and, because we are working with control systems, the category of linear spaces and linear maps $\ca{Lin}$.

The boxes at the current moment are uninhabited. Usually, we can describe the behavior of a CPS mathematically via some equations. To assign some behavior to the boxes in the wiring diagram category $\WD$, we need to construct an algebra of behaviors $F$ that takes an empty box, say $\SmallBox{}{}{Z}$, and applies the mathematical interpretation of behavior to it, $F($\SmallBox{}{}{Z}$)$. Formally, we express this as a functor from $\WD$ to $\Cat$, which assigns to every box a category representing the kind of systems that can inhabit it. Morphisms of $\WD$ map to functors that build the category of possible inhabitants of the total composite system from system components. Detailed examples working out the assignment of behavior in the case of Moore machines~\cite{spivak:2016}, contract algebras~\cite{bakirtzis:2020a}, and dynamical systems~\cite{culbertson:2020} have been already worked out in the literature. Our security framework works in any system that such a functorial assignment represents.

The other way around, one can formally decompose a box (which now is inhabited by mathematical processes) to a particular hardware and software architecture to implement a CPS. In this paper, we use an example of a UAV, but the process is repeatable for any well-formed system algebra that can take the following form.

\begin{displaymath}
	\begin{tikzcd}[row sep=.05in]
		& \WD\ar[r, "F"] & \Cat &\\
		\scriptstyle\textrm{\color{black}inner box} \ar[dddd,phantom,"{\scriptstyle\textrm{\color{fgreen}wiring}}"]
		& X{=}(X_\inp,X_\out) \ar[r,mapsto]\ar[dddd,"{\color{fgreen} f}"'] & 
		{\color{gray}FX}\ar[dddd,"{\color{purp} F(f)}"] &
		\scriptstyle\textrm{\color{gray}subsystems category}
		\ar[dddd,phantom,"{\scriptstyle\textrm{\color{purp}composite system 
				functor}}"] \\
		&&&& \\
		&&&& \\
		&&&& \\
		\scriptstyle\textrm{\color{black}outer box} & Y{=}(Y_\inp,Y_\out) \ar[r,mapsto] & {\color{gray}FY} & \scriptstyle\textrm{\color{gray}resulting system category}
	\end{tikzcd}
\end{displaymath}

\begin{figure*}[!t]
	\centering
	\includegraphics[width=1\textwidth]{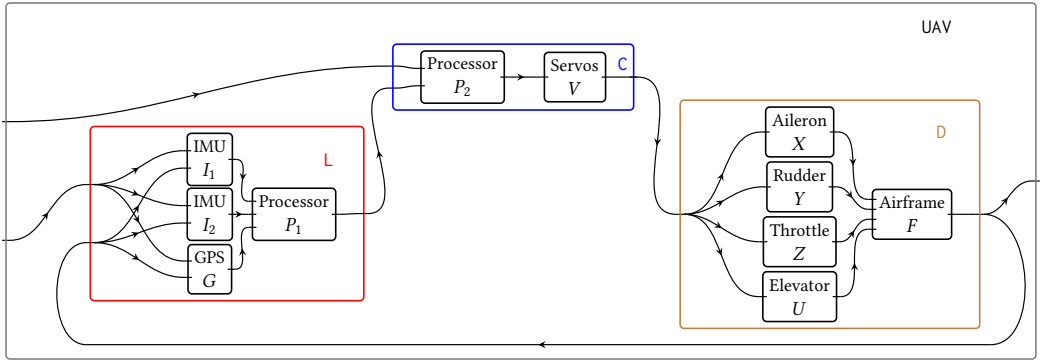}
	\caption{The hierarchical decomposition from behavior to system architecture is formally 
		contained within the slice category $\ca{C}/c$ in which there exist 
		all possible design decisions that adhere to the behavioral model. We segment here to 
		subsystems following a behavior decomposition to sensors $\mathtt{L}$, 
		controller $\mathtt{C}$, and dynamics $\mathtt{D}$. Split wires indicate function duplication, \(\Delta\).}\label{fig: UAV}
\end{figure*}

The UAV is composed of a sensor unit, denoted $\mathtt{L}$, of a controller unit, denoted $\mathtt{C}$, and of a dynamics unit, denoted $\mathtt{D}$ (Fig.~\ref{fig: UAVAtt}). We have represented the assignment of behaviors to wiring diagrams with a functor \[F: \WD \to \Cat.\] In our running example, $F(\mathtt{UAV})$ denotes the category of all the possible behaviors that we can assign to the \texttt{UAV} box. Thus, our UAV is a pair $(\mathtt{UAV}, S)$ with $S$ an object of $F(\mathtt{UAV})$ representing the particular UAV model at hand. Each of these units is itself composed of various subsystems (Fig.~\ref{ex:UAV}). Multiple possible system architectures can implement this higher-level behavior. We assume that the attacker is familiar with the general class of vehicle CPS. We focus on one possible but relatively simple system architecture for illustrative purposes. The slice category over the wiring diagram gives one model of vertical composition using category theory. 

For any category $\ca{C}$ and a fixed object $C\in\ca{C}$, the slice category $\ca{C}/C$ has as objects $\ca{C}$-morphisms with fixed target $C$, for example $f\colon A\to C, g\colon B\to C,\cdots$. The arrows in that category from some $f$ to some $g$ are $\ca{C}$-morphisms $k\colon A\to B$ between the domains, making the formed triangle
\begin{displaymath}
	\begin{tikzcd}
		A\ar[rr,"k"]\ar[dr,"f"'] && B\ar[dl,"g"] \\
		& C &
	\end{tikzcd}
\end{displaymath}
commute, namely $g\circ k=f$. Intuitively, the slice category gives us a way of breaking up morphisms into more morphisms leading to the decomposition to a particular architecture from a general understanding of the behavioral view as recorded in the wiring diagram category, $\ca{W}$.
For our running UAV example, we have multiple possible decompositions to the slice category, but we will use one that is functionally complete (Fig.~\ref{fig: UAV}).

\section{The Algebra of Attacker Actions}\label{sec: security modelling}
Using the Yoneda lemma (Appendix~\ref{sec:yoneda}) and wiring diagrams and their 
algebras (Section~\ref{sec:sas}), we will formalize attacks. This is the core of this work, and all the
concepts presented from now on are new developments.

In our effort, we embrace the perspective of the attacker. For us, an attacker is simply an actor wanting to influence or change the behavior of some given system. We do not distinguish between attacks aiming at taking control of the system (as it is common in computer hacking) and attacks seeking to influence its behavior to obtain a particular effect (as in sabotage). Our framework models attacker actions to encompass both mechanisms and domains of attacks as defined in common attack pattern enumeration and classification (CAPEC)~\cite{CAPEC}.

To describe the entire cycle of an attack, we split it into the two following main phases.
\begin{enumerate}[{Phase} 1)]
	\item \emph{Learning (exploration)}, or information gathering, is concerned with probing/eavesdropping on the system to discern its behavior. We further divide this activity into \emph{general learning}, where the attacker focuses on gathering a broad understanding of the system at hand, and \emph{specific learning}, where the attacker probes the system to understand some particular component design choices. 
\item \emph{Hijacking (exploitation)}, where the attacker, having learned enough information to understand the weak spots of a system, \emph{deploys an exploit} which takes advantage of a found architectural flaw to influence the behavior of the system in some way the attacker desires.
\end{enumerate}
\begin{example}\label{ex:UAV}
	Consider an attacker wanting to take control (or sabotage) of a UAV. The attacker starts by learning and gathering information about the target UAV (phase~1). General learning here can be the attacker trying to understand if the UAV has a GPS module on board. If there is a GPS, specific information gathering consists of understanding how the GPS module communicates with surrounding units. This general-specific learning pattern can repeat arbitrarily: the attacker could now focus on the GPS module to understand if some particular kind of integrated circuit is used in its schematic.

Once an attacker is sufficiently informed about the system, the exploit can be deployed (phase 2). In our example, this may be rewriting the firmware of the GPS module \emph{over the air} or supposing that the attacker has physical access to the UAV, manually rewiring the module, or replacing some integrated circuit in it.
\end{example}
One fundamental assumption is that we do not precisely model how a given exploit is developed but only how it is administered and provide a compositional recipe to describe how this change of behavior propagates to the whole system. In practice, this means postulating that the attacker already has access to a knowledge database of tools made to take advantage of a given structural flaw in a given (sub)system. This postulate is not unthinkable; an attacker can hijack a system without personally creating the exploit. For example,  \emph{zero day exploits} -- that is, exploits of a given flaw that is still unknown to the public, including the target manufacturer -- are common in hacking and can be commonly bought over the web~\cite{ablon:2014}. Another fundamental assumption of our model is that it forms an algebra in the wiring diagram category. Again, we do not consider this restrictive for reasons already pointed out by Bakirtzis et al.~\cite{bakirtzis:2020}. 
\subsection{Phase 1 -- Learning (Exploration)}
First, we model general learning. Assuming an attacker's perspective, we want to model what an attacker does to understand the kind of system they want to attack. In practice, this includes operations such as scanning a computer for open ports, investigating the firewall policies,  finding out what operating system is running on the system, and probing a piece of hardware to obtain information about the system's integrated circuits.

Functors $\WD \xrightarrow{F} \Cat$, $\WD$-algebras, represent wiring diagrams together with semantics linking any diagram to the category describing its possible behaviors. A particular wiring diagram is just an object $X$ in $\WD$, $FX$ is a category: objects model general behavior assignments for $X$, while morphisms take mappings between them that preserve properties we care about. When focusing on a particular system, we are fixing a wiring diagram $X$ and one of the many possible behaviors in the category $FX$. Hence,
\begin{center}
	\emph{a {system} is a pair $(X,S)$, with $X$ 
		an object\\ of $\WD$ and $S$ an object of $FX$.}
\end{center}
\begin{example}\label{ex: Moore example}
		Consider $F$ to be the $\WD$-algebra assigning each wiring diagram $X = (X_\text{in}, X_\text{out})$ to the category of Moore machines with $X_\text{in}$ and $X_\text{out}$ as input and output alphabets, respectively, as worked out in detail in~\cite{spivak:2016}. In this setting, a system $(X,S)$ is given by a wiring diagram $X=(X_\text{in}, X_\text{out})$ and a chosen Moore machine having $X_\text{in}$ as input alphabet and $X_\text{out}$ as output alphabet, respectively.
\end{example}
We suppose we have access to $X$ (this amounts to saying that we can distinguish the type of inputs and outputs that our system has) and to $FX$ (we know what kind of system we are dealing with), but not to $S$ (we do not know the specific behavior of the system at hand). The first goal of the attacker is to infer $S$ to the degree that an attack is viable.

First things first, we select a subset of the objects of $FX$, denoted $K_{FX}$ (from ``known''), representing the systems that the attacker knows or is familiar with. Notice that $K_{FX}$ should not, in general, lift to a functor $\WD \to \Cat$, since we do not assume the attacker knowledge to be compositional. For the same reason, given that there will be an injection $K_{FX} \hookrightarrow FX$ representing how the systems known by the attacker embed in the bigger universe of the systems, we do not necessarily assume the attacker to know this embedding.
\begin{definition}\label{def:F}
	Given a $\WD$-algebra $\WD \xrightarrow{F} \Cat$ and an object $X$ in $\WD$, a
	\emph{knowledge database for $F,X$} is a subset $K_{FX}$ of the objects of $FX$.
\end{definition}
		An example knowledge database for $F, X$ is just a set of Moore machines (Example~\ref{ex: Moore example}) having $X=(X_\text{in}, X_\text{out})$ as input/output alphabets, respectively.
Next, we consider functors $FX \xrightarrow[]{\Test} \Set$. These are 
interpreted as \emph{tests}, or \emph{probes}.
\begin{itemize}
	\item Given $S$ in $FX$, $\Test S$ represents the information we get in probing $S$ with a test $\Test$. For instance, $S$ may represent a machine on a network, while $\Test S$ could represent the output one gets by running \texttt{nmap} on $S$.
	
	\item If $S \xrightarrow[]{f} S'$ is a morphism of $FX$, then $\Test f$ is a way to transform the information in $\Test A$ to information in $\Test B$. Our tests are well suited to detect the properties we care about preserved by morphisms of $FX$. 
	
	\item In this interpretation, functoriality holds on the nose. Transforming a system by ``doing nothing'' (identity morphism) should give the same test outcome (functor identity law). Moreover, composing transformations should amount to composing outcomes of the testing.
\end{itemize}
We package all this information as follows.
\begin{definition}
	Given a $\WD$-algebra $\WD \xrightarrow{F} \Cat$ and an object $X$ in $\WD$, a
	\emph{test for $F,X$} is a functor $FX \to \Set$.
\end{definition}
A possible test is a functor mapping any Moore machine (Example~\ref{ex: Moore example}) to its set of states. This test is a functor because of how morphisms between Moore machines are defined; see, for instance, Schultz et al.~\cite{spivak:2016}. Again, the attacker does not have access to $S$ but has access to $\Theta S$ for some tests $\Theta$. The tests represent the ability of the attacker to perform tests on the system. These tests describe an ongoing reconnaissance mission, which often is the step that takes the longest time and resources of the attacker. The goal of the attacker is to prove in some sense that $S \simeq S'$, for some $S'$ in $K_{FX}$. The system $S$ is an instance of a system $S'$ the attacker is familiar with. Given our assumption, we can then postulate that the attacker knows an exploit for $S'$ to move to phase 2. 

We make the assumption that, for any $S' \in K_{FX}$, the attacker has access 
to $S'$. This assumption is natural, since $S'$ is by definition in the knowledge base of 
the attacker. In particular, we assume that the attacker is 
able to perform \emph{any} test to \emph{any} known system, hence,
\begin{center}
	\emph{for any $S' \in K_{FX}$ and $FX \xrightarrow{\Test} \Set$, \\the attacker has access to $\Test S'$.}
\end{center}
We extend the example of Moore machines (Example~\ref{ex: Moore example}) to include a knowledge database and security tests. The attacker, therefore, knows the set of states of all Moore machines residing in their knowledge database. The Yoneda lemma says that if $\Test S \simeq \Test S'$ for all $\Test$, then $S \simeq S'$.  In our compositional framework, let us interpret what this means, considering some corner cases.
\begin{itemize}
	\item Suppose that for some object $S$ in $FX$ there is an object $S'$ in $K_{FX}$ such that $S \simeq S'$. If the attacker has access to $\Test S$ for any $\Test$, then the attacker will be able to conclude $S \simeq S'$ from $\Test S \simeq \Test S'$. That is, if the attacker is free to perform any form of testing and possesses a vast knowledge database, then $S$ can be determined with absolute precision.
	\item If the attacker has access to any $\Test$, but there is no $S'$ in $K_{FX}$ such that $S \simeq S'$, then the attacker will not be able to conclude $S \simeq S'$. Tests can be arbitrarily precise, but the attacker cannot interpret them. 
	\item If there is an object $S'$ in $K_{FX}$ such that $S \simeq S'$, but the attacker has no access to all $\Test$, then it will not be able to conclude \emph{with certainty} that $S \simeq S'$, because the Yoneda lemma does not hold in this setting. Still, after performing enough tests, the attacker may be prone to \emph{infer} that $S \simeq S'$ if $\Theta S \simeq \Theta S'$ for \emph{enough} $\Theta$ ran. This inference comes with uncertainty, making information gathering more of an art than a science.
\end{itemize}
The Yoneda lemma provides a formal justification for the insufficiency of extensively testing a system to characterize its behavior adequately. We call this heuristic \emph{Yoneda reasoning}.

Some tests are more informative than others. For instance, the ``terminal test'' $\Theta$ sending any $X$ to the singleton set $\{ *\}$ is maximally uninformative: the result of this test is the same for any system. On the contrary, a functor that is \emph{injective on objects} lifts to a test that yields the conclusion $X = Y$ from $FX = FY$. Further formalizing the possible spectrum of tests, hopefully weighing them with probability distributions to model their reliability, is an ongoing direction of future work. We suppose that the attacker pinned down the target system $S$ with some precision. The next step of an attack is harvesting information about the architectural design choices implementing the system. After we know how $S$ works, we need to determine what $S$ is made of.

Previously, we modeled a system of an object $X$ in $\WD$ together with an object $S$ of $FX$, for some $\WD$-algebra $F: \WD \to \Cat$. Now we consider the \emph{category of architectural choices for $X$}, that is, the slice category $\WD / X$. Objects of this category are morphisms $\bigotimes_i X_i \xrightarrow{\phi} X$, while morphisms are morphisms of wiring diagrams making the following triangle commute. %
\begin{equation*}
	\scalebox{1}{
		\begin{tikzpicture}
			\node (0b) at (2,-2) {$X$};
			\node (1a) at (0,0) {$\bigotimes_i X_i$};
			\node (1b) at (4,0) {$\bigotimes_j X_j$};
			
			\draw[->] (1a) to node[above]{$f$} (1b);
			\draw[->] (1a) to node[below left]{$\phi$} (0b);
			\draw[->] (1b) to node[below right]{$\psi$} (0b);
	\end{tikzpicture}}
\end{equation*}
Architectural choices form a category, so we can repeat the reasoning in the last section using $\WD / X$ as the category we probe. Again using Yoneda, the attacker can ascertain that a given system $(X, S)$ is made of subsystems $(X_i, S_i)$, tensored and wired together by $\phi$. At this stage, the attacker still does not know anything about the $S_i$, so the process must repeat cyclically.

In practice, tests will not have to be materially re-run on every $FX_i$: the attacker likely has only access to $\Theta S$ -- every $(X_i, S_i)$ being a subsystem of $(X, S)$ that may not necessarily be exposed to external testing. Nevertheless, it will always be the case that $\Theta(F\bigotimes_i X_i) \xrightarrow{\Theta F\phi}\Theta FX$, meaning that the outputs of tests over every $S_i$ will have to be reconstructed from tests over $S$. This reconstruction adds another layer of uncertainty for the attacker, who has to devise tests for which the mapping $\Theta F \phi$ acts as transparently as possible. Again, this backs up intuition: going back to Example~\ref{ex:UAV}, if some system $(X, S)$ comprises a subsystem $(X_i, S_i)$ (say, a GPS module), then we could devise a test on $FX$ such that in $\Theta S$ the behavior of the subsystem $S_i$ is made apparent. Similarly, when running \texttt{nmap} on a system, we can get extra information about which services are running behind which port, for example, \texttt{nginx} behind port \texttt{80}. By probing the composite system, the attacker gets information about its subsystems.

Summarizing, phase 1 is modeled as follows.
\begin{enumerate}
	\item The attacker uses tests on $FX$ and Yoneda-reasoning to find 
	the system $S$ representing the semantics of $X$.
	\item The attacker uses tests on $\WD / X$ and Yoneda-reasoning to 
	find the wiring $\bigotimes_i X_i \xrightarrow{\phi} X$ representing the implementation of $X$.
	\item The attacker repeats step $1$ on any $F X_i$ of interest to 
	find the precise behavior of the subsystem marked with $X_i$.
	\item The attacker repeats step $2$ on $\WD / X_i$ to obtain more 
	information about the subsystems making up $X_i$.
	\item These steps iterate cyclically until the attacker has gathered enough information to exploit the system.
\end{enumerate}
\subsection{Phase 2 -- Hijacking (Exploitation)}
Now suppose that the attacker has a good grasp of the system behavior and architecture and model the last step, in which the system is hijacked and exploited.
We distinguish between two main kinds of attacks.
\begin{enumerate}[{Type} 1)]
	\item \emph{Rewriting attacks} change the behavior of a (sub)system. Practical examples of this are, for instance, exploiting a vulnerability in a WiFi card to rewrite its firmware and using this change of behavior to progress towards obtaining administrative privileges over the whole machine.
	\item \emph{Rewiring attacks} modify the way subsystems communicate with each other. 
\end{enumerate}
\begin{definition}[Rewriting Attack]\label{def: rewriting attack}
	Given a $\WD$-algebra $F$, systems $(X_1, S_1), \dots, (X_n, S_n)$ and a morphism $X_1 \otimes \cdots \otimes X_n \xrightarrow{w} X$, a \emph{rewriting attack} is a morphism $S_i \xrightarrow{h} S'_i$ of $FX_i$ for some $i$. The resulting system after the attack is given by the couple 
	\begin{equation*}
		\left(X,Fw(S_1, \dots, S'_i, \dots, S_n)\right) = \left(X, (Fw(\Id{S_1}, \dots, h, \dots, \Id{S_n}))(S_1, \dots, S_n)\right).
	\end{equation*}
\end{definition}

\begin{figure}[!t]
	\centering
	\includegraphics[width=.5\linewidth]{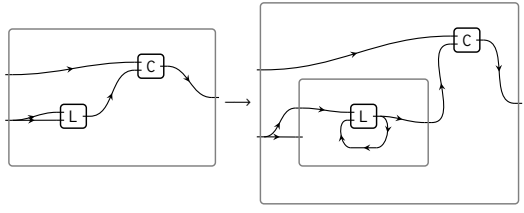}
	\caption{A rewiring attack is modelled using endomorphisms of wiring diagrams.}\label{fig: rewiring}
\end{figure}

In a rewriting attack, we do not change the possible behaviors assigned to wiring diagrams; the functor $F$ stays fixed. The pairs $(X_i, S_i)$ represent all the subsystems of our whole system, that can be evaluated as $(X,Fw(S_1, \dots, S_n))$. Here, the first component of the couple represents the shape of the box corresponding to the composite system. To pinpoint with precision the box inhabitant, we use the wiring diagram $w$, lift it to a functor between behaviors $Fw$, and evaluate this functor on the subsystems at hand, $S_1, \dots, S_n$. Then, a rewriting attack is nothing more than a way, call it $h$, to morph one of these $S_i$ into an $S'_i$. The result of swapping $S_i$ for $S'_i$ can be evaluated using the functoriality of $Fw$ and the morphism $h$. For Moore machines (Example~\ref{ex: Moore example}), a set of pairs $(X_1, S_1), \dots, (X_n, S_n)$ now represent $n$ subsystems, each one with a fixed choice of a Moore machine -- of the right kind, since each $S_i \in FX_i$ -- inhabiting it. Each of these machines will have its own set of states and state-transition function. The resulting Moore machine inhabiting $X$ will be $(X,Fw(S_1, \dots, S_n))$. 
	
	In the mapping $S_i \xrightarrow{h} S'_i$, the object $S'_i \in FX_i$ is just a Moore machine that has the same input/output alphabets of $S_i$, but possibly a different set of states or a different state-transition function. By replacing $S_i$ with $S'_i$ is our rewriting attack, we are swapping the Moore machine $S_i$ inhabiting the box $X_i$ with $S'_i$. The effect of this swap will reverberate on the overall system, and can be calculated using $Fw$. The Moore machine inhabiting $X$ is now $\left(X,Fw(S_1, \dots, S'_1, \dots, S_n)\right)$. Because of compositionality, the morphism $S_i \to S'_i$ between submachines lifts to a morphism $Fw(S_1, \dots, S_1, \dots, S_n) \to Fw(S_1, \dots, S'_1, \dots, S_n)$ between the composed machines.
	
	In our definition of rewriting attack, we keep the functor $F$ fixed, meaning that the category of behaviors to which we map each wiring diagram stays the same. We could obtain more of a granular definition of attack by considering, in addition to the elements considered in Definition~\ref{def: rewriting attack}, also monoidal natural transformations between $\WD$-algebras. This amounts to having a procedure to completely replace the \emph{kind} of systems inhabiting our boxes. For instance, the attacker could replace Moore machines (Example~\ref{ex: Moore example}) with automata of some other kind.

\begin{definition}[Rewiring Attack]
	Given a $\WD$-algebra $F$, systems $(X_1, S_1), \dots, (X_n, S_n)$ and a morphism $X_1 \otimes \cdots \otimes X_n \xrightarrow{w} X$, a \emph{rewiring attack} is a morphism $X_i \xrightarrow{h} X_i$ for some $i$. The resulting system after the attack is given by the couple 
	\begin{equation*}
		\left(X,F(w \circ (\Id{1} \otimes \dots \otimes  h \otimes \dots \otimes \Id{n} ))  (S_1, \dots, S_n)\right).
	\end{equation*}

\end{definition}

In the case of a rewiring attack, we keep $S_1, \dots, S_n$ fixed; we are not changing the specific objects inhabiting each box. What the endomorphism $h$ does is wrap a given object $S$ into new wiring before composing it with the rest of the system (Fig.~\ref{fig: rewiring}).

Both rewriting and rewiring attacks form categories. This conforms with our intuition that attacks can be performed in batches or stacked one on top of each other.

\section{Compositional Security Modeling}
To illustrate the above compositional security theory as a possible model for security tests and exploitation methods, we use a system model of a UAV from the perspective of attacker actions (Example~\ref{ex:UAV}).

Let's start with the minimum observability possible. The attacker knows that the system at hand is a couple $(\mathtt{UAV}, S)$: $\mathtt{UAV}$ with a two inputs and one output box, and $S$ is an object of $F(\mathtt{UAV})$, where $F: \WD \to \Cat$ maps system to the categories of behaviors relevant in our example. The first step of the attack is gathering information about $S$. 
\begin{figure}[!t]
	\centering
	\includegraphics[width=.5\linewidth]{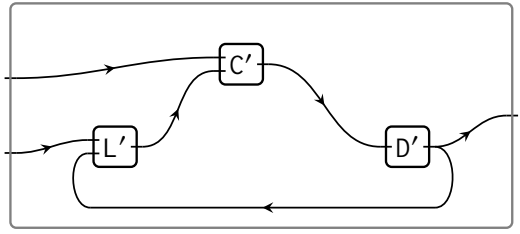}
	\caption{The attacker's understanding of the system after the first cycle of learning.}\label{fig: UAVAttt}
\end{figure}

The attacker uses Yoneda reasoning to infer $S$ in general information gathering. The attacker must be able to perform a set of tests on the system $(\mathtt{UAV}, S)$. Any of such tests is a functor \[\Theta: F(\mathtt{UAV}) \to \Set\] and the result of the test $\Theta$ applied to $S$ is denoted $\Theta S$ (in category theory practice, it is customary to avoid parentheses whenever possible). In our particular case, $\Theta$ may be a test that analyzes the aerodynamics of the UAV during flight. The more informative $\Theta$ is, and the bigger the number of $\Theta$'s the attacker can access, the more likely it will infer $S$. If the attacker finds that $S \simeq S'$ for some $S'$ in their knowledge database, $K_{F(\mathtt{UAV})}$, then the attacker will know how the UAV behaves. 

In practice, it is doubtful for the attacker to have access to \emph{every} test $\Theta$. As we mention above (Section~\ref{sec: security modelling}), this entails that the attacker won't be able to infer with certainty that $S \simeq S'$: most likely, the attacker will be \emph{prone to infer $S \simeq S'$ with a certain degree of confidence}. As such, the outcome of testing is probabilistic more than deterministic. %
\begin{figure}[!ht]
	\includegraphics[width=.7\linewidth]{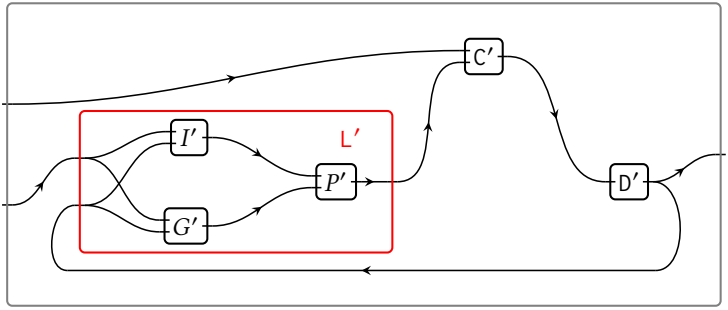}
	\caption{The architecture of the sensory system as understood by the attacker, which is in reality 
		erroneous but behaviorally equivalent. The attacker 
		found that there is one IMU (when in reality there are two) and a GPS.}\label{fig: UAVAtt2}
\end{figure}
Assuming that the attacker inferred $S \simeq S'$ for some 
system $S'$ in their knowledge database, the particular design choices making up the UAV
are still unknown to them.
Applying Yoneda reasoning again to the category $\WD / \mathtt{UAV}$, they
may be able to infer some of these design choices. For instance, it could be possible to
infer an initial understanding of what the UAV is composed of (Fig.~\ref{fig: UAVAttt}).
\begin{figure}[!t]
	\includegraphics[width=.7\linewidth]{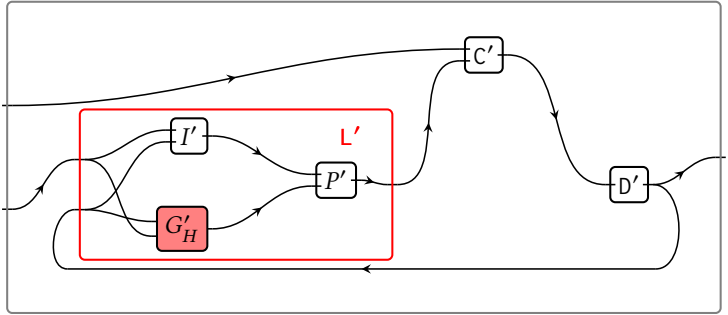}
	\caption{The compromised UAV, with GPS firmware hacked and its input wires swapped.}\label{fig: UAVAtt3}
\end{figure}
The attacker sees the system as decomposed into boxes $\mathtt{L'}$, $\mathtt{C'}$ and $\mathtt{D'}$, which will be behaviorally equivalent to $\mathtt{L}$, $\mathtt{C}$ and $\mathtt{D}$, of Fig.~\ref{fig: UAVAtt}, respectively. The inner workings of such boxes are still unknown to the attacker, which now focuses on $\mathtt{L'}$. This amounts to repeating the same cycle of Yoneda reasoning, focusing on tests that target $\mathtt{L}$ in particular (Section~\ref{sec: security modelling}). After running these tests, the attacker sees a first approximation of the UAV (Fig.~\ref{fig: UAVAtt2}). 

The initial understanding of the attacker is slightly erroneous since the two separated IMU units in $\mathtt{UAV}$ conflate into one. Still, the two wiring diagrams are behaviorally equivalent. This reflects the fact that, on the one hand, Yoneda reasoning is probabilistic, and on the other, identification of the system happens only up to behavioral equivalence.

Now, suppose the attacker decides to do two things: rewriting the firmware of the GPS module $\mathtt{G}$ and swapping its feedback inputs. The first is a rewriting attack, and the second is a rewiring attack (Section~\ref{sec: security modelling}).

The GPS module is a system $(\mathtt{G}',G')$, where $\mathtt{G}'$ is a box with two inputs and one output, while $G'$ is an object of $F\mathtt{G}'$. The rewriting attack is represented by a morphism $G' \to G'_H$ ($H$ stands for ``hacked'') in the category $F\mathtt{G}'$. Denote with 
\begin{equation*}
	\mathtt{I}' \otimes \mathtt{G}' \otimes \mathtt{P}' \otimes \mathtt{C}' \otimes \mathtt{D}' \xrightarrow{w \otimes \Id{\mathtt{C}' \otimes \mathtt{D}'}} \mathtt{L}' \otimes \mathtt{C}' \otimes \mathtt{D}' \xrightarrow{v} \mathtt{UAV}
\end{equation*}
the morphism putting together the system (Fig.~\ref{fig: UAVAtt2}). Then we have that 
\begin{equation*}
	F(v \circ (w \otimes \Id{\mathtt{C}' \otimes \mathtt{D}'}))(I',G',P',C',D') = S'
\end{equation*}
Where $I'$ is the system inhabiting $\mathtt{I}'$, $G'$ inhabits $\mathtt{G}'$ and so on, with $S'$ behaviorally equivalent to the system $S$ inhabiting $\mathtt{UAV}$.
The behavior of the UAV after the rewriting attack will be equivalent to 
\begin{equation*}
	F(v \circ (w \otimes \Id{\mathtt{C}' \otimes \mathtt{D}'}))(I',G'_H,P',C',D').
\end{equation*}

The rewiring attack, instead, is a wiring diagram $\mathtt{G}' \xrightarrow{h} \mathtt{G}'$. The following information specifies this.
\noindent\begin{minipage}{.5\linewidth}
	\begin{align*}
		\mathtt{G}_{in} \times \mathtt{G}_{out} &\xrightarrow{h_{in}} \mathtt{G}_{in}\\
		((g_1,g_2),g_3) & \xmapsto{h_{in}} (g_2,g_1)
	\end{align*}
\end{minipage}%
\begin{minipage}{.5\linewidth}
	\begin{align*}
		\mathtt{G}_{out} &\xrightarrow{h_{out}} \mathtt{G}_{out}\\
		g &\xmapsto{h_{out}} g
	\end{align*}
\end{minipage}

\smallskip
The resulting system is obtained by considering the wiring diagram
\begin{equation*}
	\mathtt{I}' \otimes \mathtt{G}' \otimes \mathtt{P}' \otimes \mathtt{C}' \otimes \mathtt{D}' \xrightarrow{\Id{\mathtt{I}'} \otimes h \otimes \Id{\mathtt{P}' \otimes \mathtt{C}' \otimes \mathtt{D}'}} 
	\mathtt{I}' \otimes \mathtt{G}' \otimes \mathtt{P}' \otimes \mathtt{C}' \otimes \mathtt{D}' \xrightarrow{w \otimes \Id{\mathtt{C}' \otimes \mathtt{D}'}} \mathtt{L}' \otimes \mathtt{C}' \otimes \mathtt{D}' \xrightarrow{v} \mathtt{UAV}
\end{equation*}
and evaluating 
\begin{equation*}
	F(v \circ (w \otimes \Id{\mathtt{C}' \otimes \mathtt{D}'}) \circ (\Id{\mathtt{I}'} \otimes h \otimes \Id{\mathtt{P}' \otimes \mathtt{C}' \otimes \mathtt{D}'}))(I',G'_H,P',C',D').
\end{equation*}
\begin{figure*}[!t]
	\centering
	\includegraphics[width=1\textwidth]{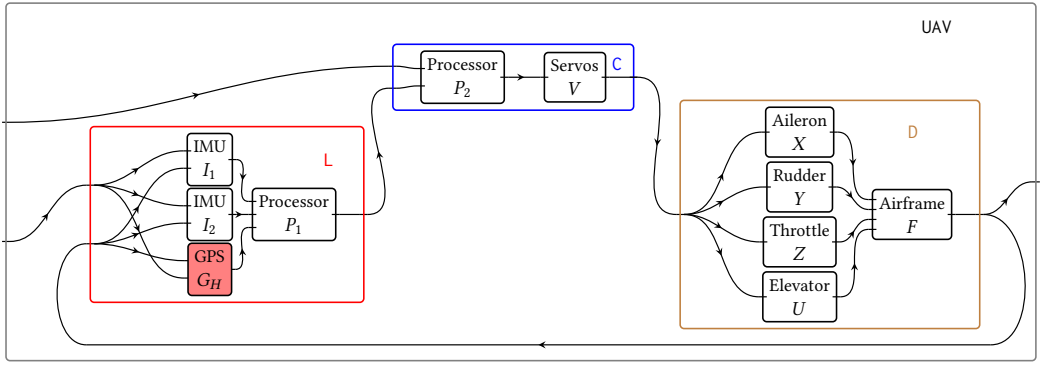}
	\caption{The UAV (Fig.~\ref{fig: UAV}), after the attack.}\label{fig: UAVAtt4}
\end{figure*}
Which is the resulting system inhabiting $\mathtt{UAV}$ after applying both attacks (Fig.~\ref{fig: UAVAtt3}). The attacker presumes how the system will behave after the exploit, assuming they have a correct system profile from phase 1. Since phase 1 has a margin for error, the attacker can re-probe the system to assure that the perceived behavior is compatible with reality. This further round of testing is necessary to assert with confidence that the exploit has been deployed correctly.

In this example, we postulated that the attacker was indeed able to gather information about the UAV correctly. Formally, we expressed this by stating what we consider to be the actual UAV (Fig.~\ref{fig: UAV}) and the attacker's understanding of what the UAV is (Fig.~\ref{fig: UAVAtt2}) are behaviorally equivalent. Because functors preserve isomorphisms, we can describe how the attack impacts the actual UAV (Fig.~\ref{fig: UAVAtt4}).

We now describe several other possible attacks: one consists in feeding a counterfeit GPS signal to the UAV to compromise it. This attack is documented ``CAPEC-627: Counterfeit GPS Signals'' and is considered difficult to realize. The wiring diagrams formalism gives us an idea of why: feeding a counterfeit GPS signal does not involve modifying the GPS module. What changes is the information traveling on the GPS wires, which communicate with the outside world. 
\begin{figure}[ht]%
	\includegraphics[width=.8\linewidth]{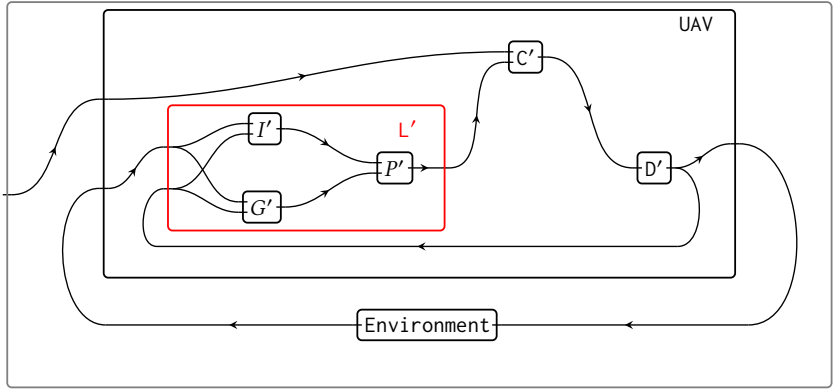}
	\caption{Feeding counterfeit GPS signals to the UAV hijacks  $\mathtt{Environment}$.}\label{fig: UAVAttEnv}
\end{figure}
So, to understand this attack properly, one needs to model how the UAV interacts with the environment it is in (Fig.~\ref{fig: UAVAttEnv}). Here, by $\mathtt{Environment}$ we mean a process that, given the UAV position in space and time, returns the data sensed by the IMU and GPS units. We see that spoofing a GPS signal does not amount to intervening on the UAV but on the environment itself. Attackers rarely can control the environment within a region of space and time -- radio waves from the GPS satellites in this particular case -- that is big enough to influence the behavior of the single UAV.
\begin{figure}[ht]
	\includegraphics[width=.8\linewidth]{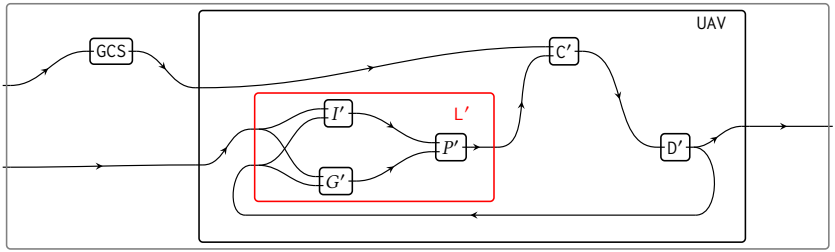}
	\caption{Social engineering attacks may hijack the ground control station ($\mathtt{GCS}$).}\label{fig: UAVAttSocial}
\end{figure}

To conclude, we present another possible attack, performed using \emph{social engineering}. As with the previous example, social engineering does not exploit the UAV itself but instead takes advantage of the human factor. Examples of this may include bribing whoever programs the UAV goals or making the control tower believe that a given order has been officially issued from whoever is in command, for example, as defined in ``CAPEC-137: Parameter Injection''.

As in the previous case, the behavior of the UAV as a wiring diagram is unchanged. Instead, what changes is the information traveling on the first input wire of the $\mathtt{UAV}$ box. From our perspective, this requires again to put the UAV into context (Fig.~\ref{fig: UAVAttSocial}). An attack based on social engineering will consist in rewriting the box $\mathtt{GCS}$, which abstracts away a possible ground control station.

The categorical semantics of CPS security modeling for the UAV can be implemented algorithmically (Listing~\ref{lst:theta} \&~\ref{lst:eta}).

\begin{lstlisting}[caption=Modeling attacker learning,label={lst:theta}]
	-- Define category of wiring diagrams
	*$\WD$* : Category
	*$\WD$*  = Definition *\ref{def:w}*
	
	-- Define functor for UAV modeling
	F : Functor *$\WD$* *$\rightarrow$* *$\Cat$*
	F = assignment of UAV behavior (linear time invariant system) to boxes
	
	-- Model UAV as a 2-input 1-output *$\color{gray}\WD$*-box
	*${\mathtt{UAV}}$*  : *$\WD$*
	*${\mathtt{UAV}}$* = (2,1)
	
	-- Knowledge database
	*$K_{F(\mathtt{UAV})}$*: List *${F(\mathtt{UAV})}$*
	*$K_{F(\mathtt{UAV})}$* = attacker knowledge for systems of type UAV
	
	-- Compare tests with target S
	CompareTests : (Functor *${F(\mathtt{UAV})}$* *$\rightarrow$* *$\Set$*) *$\rightarrow$* *${F(\mathtt{UAV})}$* *$\rightarrow$* Bool
	CompareTests *$\Theta$* S' = *$\Theta(K_{F(\mathtt{UAV})}(S')) \simeq \Theta(S)$*
	
	-- Yoneda reasoning
	for each *$\Theta$* : Functor *${F(\mathtt{UAV})}$* *$\rightarrow$* *$\Set$*
	filter (CompareTests *$\Theta$*) K
	
	-- Running security tests reveals the following boxes
	*${\mathtt{L'}}$*, *${\mathtt{C'}}$* , *${\mathtt{D'}}$*  : *$\WD$*
	*${\mathtt{L'}}$*, *${\mathtt{C' }}$* = (2,1) 
	*${\mathtt{D'}}$* = (1,1)
	
	diagram : Morphism *$\WD$* *$({\mathtt{L' \otimes C' \otimes D'}}) \rightarrow {\mathtt{UAV}}$*
	diagram = (in, out)
	where
	in : Morphism *$\Set$* *${\mathtt{UAV_{in}}} \times (\mathtt{L'}_{out} \times \mathtt{C'}_{out} \times \mathtt{D'}_{out}) \rightarrow (\mathtt{L'}_{in} \times \mathtt{C'}_{in} \times \mathtt{D'}_{in})$*
	in u1 u2 l c d = (u2, d, u1, c)
	out : Morphism *$\Set$* *$(\mathtt{L'}_{out} \times \mathtt{C'}_{out} \times \mathtt{D'}_{out}) \rightarrow \mathtt{UAV}_{out}$*
	out l c d = d
	
	
\end{lstlisting}

\begin{lstlisting}[caption=Modeling hijacking,label={lst:eta}]
	-- By iterating learning further decompose *$\mathtt{L'}$*
	*${\mathtt{I'}}$*, *${\mathtt{G'}}$* , *${\mathtt{P'}}$*  : *$\WD$*
	*${\mathtt{I'}}$*, *${\mathtt{G'}}$* , *${\mathtt{P'}}$* = (2,1)
	
	Ldiagram : Morphism *$\WD$* *$({\mathtt{I' \otimes G' \otimes P'}}) \rightarrow {\mathtt{L'}}$*
	Ldiagram = (in, out)
	where
	in : Morphism *$\Set$* *$({\mathtt{L'}_{in}} \times (\mathtt{I'}_{out} \times \mathtt{G'}_{out} \times \mathtt{P'}_{out}) \rightarrow (\mathtt{I'}_{in} \times \mathtt{G'}_{in} \times \mathtt{P'}_{in}))$*
	in l1 l2 i g p = (l1, l2, l1, l2, i, g)
	out : Morphism *$\Set$* *$(\mathtt{I'}_{out} \times \mathtt{G'}_{out} \times \mathtt{P'}_{out}) \rightarrow \mathtt{L'}_{out}$*
	out i g p = p
	
	-- Rewriting attack
	*$\eta$* : NatTrans (Functor *$\WD$* *$\rightarrow$* *$\Cat$*) *$\rightarrow$* (Functor *$\WD$* *$\rightarrow$* *$\Cat$*)
	
	-- *$\color{gray}\eta$* is the identity on everything but G'
	*$\eta$* *$\mathtt{G'}$* : Morphism F *$\mathtt{G'}$* *$\rightarrow$* F *$\mathtt{G'}$*
	*$\eta$* *$\mathtt{G'}$* = firmware rewriting
	
	-- Rewiring attack 
	Lattack : Morphism *$\WD$* *$({\mathtt{I' \otimes G' \otimes P'}}) \rightarrow {\mathtt{L'}}$*
	Lattack = (in, out)
	where
	in : Morphism *$\Set$* *$(\mathtt{L'}_{in} \times (\mathtt{I'}_{out} \times \mathtt{G'}_{out} \times \mathtt{P'}_{out}) \rightarrow (\mathtt{I'}_{in} \times \mathtt{G'}_{in} \times \mathtt{P'}_{in})$*
	in l1 l2 i g p= (l1, l2, l1, 0, i, g)
	out : Morphism *$\Set$* *$(\mathtt{I'}_{out} \times \mathtt{G'}_{out} \times \mathtt{P'}_{out}) \rightarrow \mathtt{L'}_{out}$*
	out i g p = p
	
	Rewiring : Functor *$\WD$* *$\rightarrow$* *$\WD$*
	Rewiring Ldiagram = Lattack
	
	-- The modified behavior of the hijacked UAV
	behavior : F(*${\mathtt{UAV}}$*)
	behavior = *$\eta$* *${\mathtt{UAV}}$* (F(Rewiring (*${\mathtt{UAV}}$*))) (S')
\end{lstlisting}

\vspace{1em}
\noindent
\textsc{limitations} \quad Based on how we defined Yoneda reasoning, we identify several limitations. These limitations can be overcome by enriching over metric spaces, which we will also discuss. The main point of this paper is to set a solid theoretical footprint of category theory and the diagrammatic reasoning that emerges in the application of securing CPS. Making the results probabilistically concrete is a potential future topic that can be based on the above formal methods.

One such limitation can be inspected from the resulting algorithm (Listing~\ref{lst:theta}). As output, we may have that Yoneda reasoning returns no result (the list being filtered from \texttt{K} is empty, meaning that the attacker does not have entries in the knowledge database that adequately model the target system). However, we may also have that it returns more than one -- the list being filtered from \texttt{K} having more than one element, meaning that the test performed was not fine-grained enough to pinpoint the target system with deterministic accuracy. This discrepancy is mainly due to the nature of the tests performed; some tests are more informative than others. Sending any system in $F(\mathtt{UAV})$ to the one-element set defines a functor to $\Set$ and hence a valid test, which is though maximally uninformative since the test outcome will be the same on all systems. Contrastly, any injective-on-objects functor $\Theta$ allows us to conclude that $\Theta S = \Theta S'$ implies $S = S'$, and is maximally granular. As we presented it, the formalism cannot express which subset of the tests allows us to individuate the target system unambiguously. 

Indeed, there are tests with different degrees of expressiveness, and the Yoneda lemma does not account for this; we can conclude $S \simeq S'$ using Yoneda lemma if and only if $S$ and $S'$ agree on \emph{all} tests, including the maximally useless ones. These different results are why we speak of Yoneda reasoning as a heuristic and not as a deterministic procedure.

Looking at things more abstractly, the reason for this shortcoming lies in the fact that in our definition of the category, we considered homsets to be sets; that is, we speak of the \emph{set} $\Hom{\CategoryC}{A}{B}$ of all possible morphisms from $A$ to $B$ in some category $\CategoryC$. Sets have very little structure, and in such an environment, we cannot formulate the Yoneda lemma to be more expressive. 

In a probabilistic setting, what we would like to have is a version of Yoneda reasoning that gives an \emph{interval of confidence} relating $\Theta S \simeq \Theta S'$ and $S \simeq S'$ for any possible test $\Theta$. In other words, we want to attach to each $\Theta$ a measure of how informative $\Theta$ is in our context.

One possible solution is to resort to \emph{enriched category theory}, which is a generalization of category theory where homsets can have more structure. In particular, we can reformulate our theory using categories enriched over metric spaces. Categories enriched over metric spaces give a natural way of talking about \emph{distances} between sets, and this can be used to define a measure on the tests we can perform. In the context of enriched category theory, the Yoneda lemma can be reformulated in what is informally known as \emph{ninja Yoneda lemma}~\cite{Loregian2019}, which takes into account this additional structure. We can use this to define a version of Yoneda lemma that has a notion of confidence in the tests we perform over the CPS model and, therefore, have some granularity of what it means for two systems to be behaviorally equivalent under \emph{some} (informative) tests. \vspace{1em}

\noindent \textsc{benefits} \quad While we show some limitations about the flexibility of the model, it is essential to point out that the same flexibility can be beneficial. Different formalisms can inhabit the boxes defining a system's behavior, from Petri nets to transition systems to ordinary differential equations. Developing our formalisms will allow us to speak about all these representations within one framework. For example, applications to security modeling using Petri nets~\cite{lesi:2020} is currently congruent with research in category theory and Petri nets~\cite{baez:2020, genovese:2019b, genovese:2019c, StateboxTeam2019} and could be used to make the application of the preceding formalism more concrete as a model of (mis-)behavior. Similarly, both models of security violations in automata~\cite{wang:2019} and continuous controller behavior~\cite{pajic:2014} can be represented within our framework and, therefore, allow for a plethora of analyses within one model.

However, to drill further in the possible directions of describing different types of continuous, for example, false sensor data, or discrete, for example, transitioning the system to a hazardous state, misbehaviors caused by exploitation require the first formulation of security modeling categorically and algebraically. This paper serves this purpose. Relaxing some of the unrealistic assumptions we made incorporates developing work from category theory.

Additionally, this security framework is part of an alternate paradigm of systems modeling that has its foundations in categorical modeling. In this framework, it is possible to provide formal traces of requirements, behaviors, and architectures~\cite{bakirtzis:2020a} but also describe a vast amount of dynamical systems with applications to robotics, event-based systems, and hybrid systems, to name a few~\cite{libkind:2020,fong:2019,zardini:2020,culbertson:2020}.

Finally, this paper addresses the theoretical underpinnings of security modeling in category theory. Nevertheless, the recent surge of categorical modeling languages and software, such as Catlab~\cite{halter:2020} or idris-ct~\cite{genovese:2019d} or algebraic databases~\cite{schultz:2016}, can be used to create modeling tools and security assessment methods based on the work presented in this paper practical within compositional CPS theory.

\section{Related Work}

In general, category theory is effective in describing hybrid systems~\cite{tabuada:2002,ames:2006,culbertson:2020} and more recently there has been successful work in modeling and analysis of CPS using category theory~\cite{nolan:2019,breiner:2019a,breiner:2019,bakirtzis:2020}. An important motivation for developing a categorical \emph{modeling} security framework is the theory of co-design, a way of dealing with abstraction and refinement in models, which has recently been applied categorically to robotics~\cite{zardini:2020b,zardini:2020c} and control system design~\cite{zardini:2020a}.

To the best of our knowledge, there is little work at the intersection of category theory, CPS, and security \emph{modeling}. One such work uses the categorical interpretation of databases to share threat information, but it does not propose how this paradigm improves upon graph methods for security~\cite{andrian:2017}. A line of work also uses category theory to study cryptographic functions, as illustrated, for example, by Pavlovic~\cite{pavlovic:2014}. The modeling approach presented in this paper does not develop security techniques for defenses but instead focuses on the dual of the research questions answered by cryptography. While cryptography asks how we can communicate securely, systems modeling and analysis ask how we can represent attacker actions over a behavioral and architectural model in a traceable manner, such that we can examine what mitigation strategies to implement in the design of the system. The same differentiation applies to secure design using dependent types~\cite{hennessy:2005} -- which are often formalized categorically -- and categorical data flow analysis~\cite{zhu:2014}.

On the side of attacker modeling for CPS, there is a vast area of research using graph formalisms. Examples include attack graphs~\cite{sheyner:2002} and attack profiles using graph models as shown recently by Weber et al.~\cite{weber:2020}, including particular applications to industrial control systems~\cite{ghazo:2019}. Our wiring diagram model can be thought of as a graph with extra structure, namely the added structure of the category. Therefore such methods could also be incorporated into our algebraic security tests paradigm.

From the controls or system behavior view, defenders can intercept the \emph{learning} phase of attacker actions by adding a privacy-enhancing signal into the controller~\cite{khojasteh:2020}. From purely modeling, which is more closely related to techniques from reliability and dependability, recent work aims to merge attack trees with standard design practices for embedded systems~\cite{li:2018}. These frameworks are relevant to one abstraction level, that of system behavior, and could be subsumed by the categorical formalism we present above. In the future, both the controls and modeling approaches could be improved within our framework by providing formal composition and traceability between the expected behavior and the eventual synthesized design.

The intersection of formal methods, control, and CPS security is addressed through differential dynamic logic. Differential dynamic logic has provided a plethora of verification capabilities for the continuous and discrete parts of CPS~\cite{platzer:2020}. Differential dynamic logic has also been applied for security modeling~\cite{bohrer:2018}. We see differential dynamic logic as complementary to our framework and vice versa. Categorical primitives can become part of differential dynamic logic, while dynamic logic can provide richer semantics for how the behavior of the whole system changes based on specific modeled attacks.

All in all, we perceive applications of category theory to model-based security as relatively unfledged and hope to see category theory used as effectively in security modeling as it has for programming languages~\cite{pierce:1991,stay:2013,gratzer:2020, genovese:2019d} and cryptography~\cite{datta:2005, genovese:2019a}.

\section{Conclusion}
We develop a categorical semantics for CPS security modeling that can determine that two system representations are behaviorally equivalent, provided that they agree on every test. This statement implies that it is possible to model attacker actions without giving the attacker full observability of the attack system. Additionally, we model two types of attacks on the incomplete but erroneous view of the attacker and show its impact on (what we consider to be) the existing system. These attacks can either (1) rewrite some system component or (2) rewire an input or output from or to a component. This model is beneficial for CPS. In the future, we would like to say how a particular attack can change system behavior and, therefore, potentially transition it to a hazardous state. Overall, we model how the attacker \emph{learns} about a system and how an attacker then might attempt to \emph{hijack} the system from the knowledge that they were able to gather in a formal, unified way. Finally, the categorical formalism can be considered foundational. In addition to the contributions above, it can subsume already developed formalisms for modeling attacker actions, such as attack graphs, or augmenting the information contained in the model by using security frameworks.

\begin{acks}
The authors thank D. Evans
and C. Vasilakopoulou
for constructive discussions and feedback.
\end{acks}

\bibliographystyle{ACM-Reference-Format}
\bibliography{manuscript}
\appendix
\section{Categorical Preliminaries}\label{sec: preliminaries}
\subsection{Categories}
Category theory is a framework to study patterns in mathematics. 
The fundamental definition is the one of a category.
\begin{definition}\label{def: category}
	A category $\CategoryC$ is composed of:
	\begin{itemize}
		\item A collection of \emph{objects}, denoted $\Ob{\CategoryC}$;
		\item For each pair of objects $A,B$, a collection of \emph{morphisms from $A$ to $B$}, denoted $\Hom{\CategoryC}{A}{B}$.
		A morphism $f$ in $\Hom{\CategoryC}{A}{B}$ is usually denoted as $A \xrightarrow{f} B$;
		\item For each object $A$, a morphism $A \xrightarrow{\Id{A}} A$;
		\item For each $A,B,C$ objects, an operation 
		\begin{equation*}
			\circ_{A,B,C}: \Hom{\CategoryC}{B}{C} \times \Hom{\CategoryC}{A}{B} \to \Hom{\CategoryC}{A}{C}
		\end{equation*}
		called \emph{composition}. We usually omit the subscripts and just write $g \circ f$ to denote composition. Composition
		is also denoted diagrammatically, as in $A \xrightarrow{f} B \xrightarrow{g} C$.
		\item Finally, we require the following equations to hold for each $A \xrightarrow{f} B$, $B \xrightarrow{g} C$ and $C \xrightarrow{h} D$:
		\[
		\Id{B} \circ f = f \qquad f \circ \Id{A} = f \qquad (h \circ g) \circ f = h \circ (g \circ f)
		\]
	\end{itemize}
\end{definition}
The interpretation we give to categories is the following: Objects can represent systems, states of a given system, or, in general, entities we care about. Morphisms represent transformations between these entities. The axioms amount to asking the following.
\begin{itemize}
	\item For each system $A$ there is a ``do nothing'' transformation, called $\Id{A}$.
	\item We can compose transformations whenever their domain and codomain match. This captures the idea 
	of applying transformations \emph{sequentially}, each transformation acting on the result of the previous one.
	We require this composition to be associative.
\end{itemize}
\begin{example}
	The simplest example of category in this context is $\Set$, the category
	whose objects are sets and morphisms are functions between them. 
	For each set $A$, $\Id{A}$ is the identity function from $A$ to itself;
	composition is function composition.
\end{example}
\begin{example}
 Plenty of familiar structures in mathematics can be seen as categories. For instance, a monoid can be seen as a category with only one object, call it $*$. Any element of the monoid is interpreted as a morphism $* \to *$. The identity on $*$ is the monoid unit, and composition is the monoid operation. Indeed, categories can be thought of as generalized monoids with many objects. 

Other familiar categories include groups and their homomorphisms, vector spaces and linear maps between them, topological spaces and continuous functions, and the category of states and transitions between them~\cite{diskin:2015}.
\end{example}
In general, the idea is that morphisms are transformations that preserve some properties possessed by the objects, properties that we care about. So, for instance, if we were to define a category of topological spaces, we may require morphisms to be continuous functions or homotopies. If we define a category whose objects are algebras, we may require the morphisms to be homomorphisms that preserve the relevant algebraic properties we want to study.

What happens when two objects behave exactly in the same way with respect to the properties we are interested in? This notion can be categorified as:
\begin{definition}\label{def: isomorphism}
	Given a category $\CategoryC$, a morphism $A \xrightarrow{f} B$ is called an \emph{isomorphism}
	if there is a morphism $B \xrightarrow{f^{-1}} A$ such that the following square \emph{commutes}, meaning that 
	any two paths sharing the same start and end points define the same morphism:
	\begin{equation*}
		\scalebox{1}
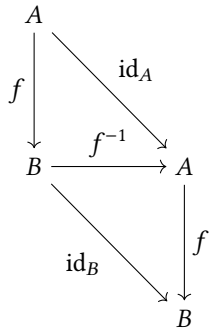
{
			\begin{tikzpicture}
				\node (0b) at (2,-2) {$B$};
				\node (1a) at (0,0) {$B$};
				\node (1b) at (2,0) {$A$};
				\node (2a) at (0,2) {$A$};
				
				\draw[->] (2a) to node[left]{$f$} (1a);
				\draw[->] (1a) to node[above]{$f^{-1}$} (1b);
				\draw[->] (1b) to node[right]{$f$} (0b);
				
				\draw[->] (2a) to node[above right]{$\Id{A}$} (1b);
				\draw[->] (1a) to node[below left]{$\Id{B}$} (0b);
		\end{tikzpicture}}
	\end{equation*}
	If there is an isomorphism $A \xrightarrow{f} B$, then we say that 
	$A$ and $B$ are \emph{isomorphic}, and write $A \simeq B$.
\end{definition}
\begin{remark}
	Importantly, two isomorphic objects in a category behave exactly in the same way
	\emph{only} with respect to the structure captured by the category. For instance, $\Reals$
	and $\Reals^2$ can be seen both as objects in $\Set$, the category of sets and functions, 
	and as objects in $\VectR$, the category of real vector spaces and linear maps between them. 
	They are isomorphic in $\Set$, since bijective functions satisfy Definition~\ref{def: isomorphism}
	and $\Reals$ and $\Reals^2$ are in bijection. Nevertheless, they are \emph{not} isomorphic
	in $\VectR$, since an isomorphism in this category has to be a \emph{linear} bijection, which in particular
	has to preserve dimension. What is happening here is that since $\VectR$ keeps track of more structure than
	what $\Set$ does, our ability to tell objects apart in $\VectR$ is \emph{finer} than in $\Set$.
\end{remark}
\subsection{Functors}
Functors are morphisms between categories. As we said in the previous section, they should then preserve 
the properties we care about when we study categories. Looking at Definition~\ref{def: category}, 
these are just identities and composition. Hence, we give the following definition.
\begin{definition}\label{def: functor}
	Given categories $\CategoryC, \CategoryD$, a functor 
	$\CategoryC \xrightarrow{F} \CategoryD$ consists of the following information:
	\begin{itemize}
		\item A mapping $\Ob{\CategoryC} \xrightarrow{F} \Ob{\CategoryD}$;
		\item For each $A,B \in \Ob{\CategoryC}$, a mapping
		\begin{equation*}
			\Hom{\CategoryC}{A}{B} \xrightarrow{F} \Hom{\CategoryD}{FA}{FB}
		\end{equation*}
		\item We moreover require the following equations to hold:
		\begin{equation*}
			F(\Id{A}) = \Id{FA} \qquad F(g \circ f) = F(g) \circ F(f)
		\end{equation*}
	\end{itemize}
\end{definition}
Functors are structure preserving maps that allow us to connect different model types
by defining the particular semantics of transformations that are necessary
to change the domain of discourse (within a particular category, say $\Set \to \Set$,
or between different categories, say $\CategoryC \to \Set$).
\begin{remark}
	In category theory practice it is customary to omit parentheses when not 
	strictly necessary. As such, we will often write $Ff$ instead of 
	$F(f)$ to denote the application of a functor $F$ to a morphism $f$.
\end{remark}
\begin{example}
	There is a functor from $\VectR$ to $\Set$ that ``forgets structure'':
	Any real vector space is mapped to its underlying set, and any linear map
	between them is mapped to its underlying function between sets.
\end{example}
\begin{example}
	As we will see shortly, a very important functor in category theory is the hom-functor: 
	Fix an object $A$ in a category $\CategoryC$. Then we can define a functor
	\begin{equation*}
		\CategoryC \xrightarrow{\Hom{\CategoryC}{A}{-}} \Set   
	\end{equation*}
	Which sends every object $B$ of $\CategoryC$ to the set of morphisms 
	$\Hom{\CategoryC}{A}{B}$. A morphism $B \xrightarrow{g} C$ is sent to the function
	\begin{equation*}
		\Hom{\CategoryC}{A}{B} \xrightarrow{\Hom{\CategoryC}{A}{g}} \Hom{\CategoryC}{A}{C}
	\end{equation*}
	Which acts by postcomposing: a morphism $A \xrightarrow{f} B$ is sent to 
	$A \xrightarrow{f} B \xrightarrow{g} C$. Functoriality follows 
	from the composition and identity axioms of $\CategoryC$.
\end{example}
Functors are also useful to supply a category with an additional operation alongside $\circ$, which
intuitively models the idea of of considering multiple objects at the same time and performing
transformations \emph{in parallel}.
\begin{definition}
	A \emph{monoidal} category $\CategoryV$ is a category that comes equipped
	with a \emph{monoidal product functor}
	\begin{displaymath}
		\CategoryV \times \CategoryV \xrightarrow{\otimes} \CategoryV
	\end{displaymath}
	which can be thought of as multiplication of objects and morphisms, or more 
	broadly as doing operations in parallel. 
	We require that, for any objects $X,Y,Z$,
	\begin{equation*}
		(X\otimes Y)\otimes Z \simeq X\otimes (Y\otimes Z)
	\end{equation*}
	Meaning that multiplying objects in any order gives isomorphic results.
	
	We also require the existence of a distinguished object $I$ of $\CategoryV$, 
	called \emph{monoidal unit}, such that 
	\begin{equation*}
		I \otimes X \simeq X \simeq X \otimes I
	\end{equation*} 
	that is, $I$ acts like an identity for this multiplication. All this data must satisfy certain 
	axioms~\cite{BraidedTensorCats}, that are beyond the scope of this paper.
\end{definition}
\begin{example}
	Widely used examples of monoidal categories include $(\Set,\times,\{\star\})$, 
	with the cartesian product of sets and the singleton,
	as well as $(\VectR,\otimes,\Reals)$, with the tensor product of vector spaces.
	Moreover $\Cat$, the \emph{category of categories and functors between them},
	admits a monoidal structure $(\Cat,\times,\ca{1})$, where tensor is defined as 
	the cartesian product of categories (similarly to that of sets), and
	the tensor unit is the category $\ca{1}$ consisting of a single object together with its identity arrow.
	
	In fact, all these are examples of \emph{symmetric} monoidal categories, 
	which come further equipped with isomorphisms
	\begin{equation*}
		X\otimes Y\simeq Y\otimes X.
	\end{equation*} 
	For example, for two sets it is $X\times Y\simeq Y\times X$ via the mapping $(x,y)\mapsto (y,x)$.  
\end{example}
Now that we defined monoidal categories, that are nothing but categories
together with some additional structure, we have to refine our notion of functor accordingly.
\begin{definition}
	A \emph{monoidal} functor between two monoidal categories 
	$(\CategoryV,\otimes_\CategoryV,I_\CategoryV) \xrightarrow{F} (\CategoryW,\otimes_\CategoryW,I_\CategoryW)$ 
	is a functor that preserves the monoidal structure in a lax sense (meaning not up to isomorphism).
	Namely, it comes equipped with morphisms 
	\begin{align*}
		F(I_\CategoryV) &\xrightarrow{\phi_0} I_\CategoryW\\
		FX\otimes_\CategoryW FY&\xrightarrow{\phi_{X,Y}} F(X\otimes_\CategoryV Y)
	\end{align*}
	with $X,Y$ ranging over the objects of $\CategoryV$, that express the relation between the image 
	of the tensor and the tensor of the images inside the target category $\CategoryW$; these adhere 
	to certain axioms \cite{BraidedTensorCats}.
\end{definition}

\subsection{Natural Transformations}
In this work we will also use the fundamental concept
of the natural transformation. 
A natural transformation models the transformation
between functors while preserving structure,
otherwise called morphism of functors.
\begin{definition}
	Given functors $\CategoryC \xrightarrow{F,G} \CategoryD$, a \emph{natural transformation} 
	$F \xRightarrow{\eta} G$ consists, for each object $A$ of $\CategoryC$, of a morphism 
	$FA \xrightarrow{\eta_A} GA$ in $\CategoryD$ such that, for each morphism $A \xrightarrow{f} B$ in $\CategoryC$,
	\begin{equation*}
		\eta_B \circ Ff = Gf \circ \eta_A.
	\end{equation*}
	This is often expressed diagrammatically by saying that the following square has to commute:
	\begin{equation*}
		\scalebox{1}{
			\begin{tikzpicture}
				\node (1a) at (0,0) {$FB$};
				\node (1b) at (2,0) {$GB$};
				\node (2a) at (0,2) {$FA$};
				\node (2b) at (2,2) {$GA$};
				
				\draw[->] (2a) to node[above]{$\eta_A$} (2b);
				\draw[->] (1a) to node[below]{$\eta_B$} (1b);
				\draw[->] (2a) to node[left]{$Ff$} (1a);
				\draw[->] (2b) to node[right]{$Gf$} (1b);
		\end{tikzpicture}}
	\end{equation*}
\end{definition}
Intuitively, the information contained in a natural transformation $F \xRightarrow{\eta} G$ 
is enough to guarantee that any commutative diagram made of images of things in 
$\CategoryC$ via $F$ can be turned into a diagram of images of things in $\CategoryC$ via 
$G$ without breaking commutativity. An example is shown in the figure below: 
The commutativity condition of $\eta$ means that it doesn't matter in which 
order we will ``walk through'' these arrows, the result will be the same.
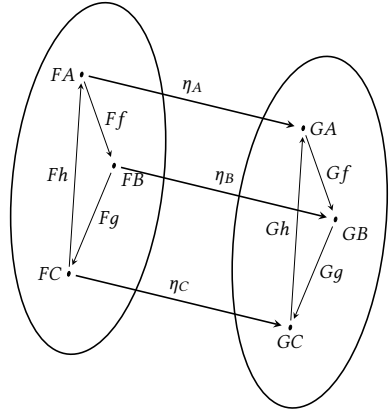
\begin{equation*}
	\scalebox{0.75}{
		\tdplotsetmaincoords{70}{125}
		\begin{tikzpicture}[tdplot_main_coords, scale=0.4]
			
			\begin{scope}[canvas is xz plane at y=0]
				\draw[thick] (0,0) ellipse (6cm and 8cm);
				
				\draw[fill] (0.5,5) circle (3pt) node[align=left, black, left] {$FA$};
				\node (f1) at (0.5,5) {};
				\draw[fill] (-2,0) circle (3pt) node[align=center, black, below right] {$FB$};
				\node (g1) at (-2,0) {};	
				\draw[fill] (1.5,-4) circle (3pt) node[align=right, black, left] {$FC$};
				\node (h1) at (1.5,-4) {};	
				
				\draw[->, >=stealth, shorten >=.03cm] (f1) to node[right, black] {$Ff$} (g1);
				\draw[->, >=stealth, shorten >=.03cm] (g1) to node[right, black] {$Fg$} (h1);
				\draw[->, >=stealth, shorten >=.03cm] (h1) to node[left, black] {$Fh$} (f1);
				
			\end{scope}
			
			\begin{scope}[canvas is xz plane at y=12]
				\draw[thick] (0,0) ellipse (6cm and 8cm);
				
				\draw[fill] (0.5,5) circle (3pt) node[align=left, black, right] {$GA$};
				\node (f2) at (0.5,5) {};
				\draw[fill] (-2,0) circle (3pt) node[align=center, black, below right] {$GB$};
				\node (g2) at (-2,0) {};	
				\draw[fill] (1.5,-4) circle (3pt) node[align=right, black, below] {$GC$};
				\node (h2) at (1.5,-4) {};	
				
				\draw[->, >=stealth, shorten >=.03cm] (f2) to node[right, black] {$Gf$} (g2);
				\draw[->, >=stealth, shorten >=.03cm] (g2) to node[right, black] {$Gg$} (h2);
				\draw[->, >=stealth, shorten >=.03cm] (h2) to node[left, black] {$Gh$} (f2);
				
				\draw[->, >=stealth, shorten >=.03cm, thick] (f1) to node[above, black]{$\eta_A$} (f2);
				\draw[->, >=stealth, shorten >=.03cm, thick] (g1) to node[above, black]{$\eta_B$} (g2);
				\draw[->, >=stealth, shorten >=.03cm, thick] (h1) to node[above, black]{$\eta_C$} (h2);
			\end{scope}
	\end{tikzpicture}}	
\end{equation*}
\subsection{The Yoneda Lemma}
\label{sec:yoneda}
The Yoneda Lemma is arguably the most important result in category theory. 
It follows from a clever observation: consider a functor $\CategoryC \xrightarrow{F} \Set$ 
and an object $A$ of $\CategoryC$. By definition, any natural transformation 
$\Hom{\CategoryC}{A}{-} \xRightarrow{\eta}  F$ has as components 
functions between \emph{sets} $\Hom{\CategoryC}{A}{B} \xrightarrow{\eta_B} FB$.
Moreover, for each $A \xrightarrow{f} B$, the following diagram must 
commute because $\eta$ is a natural transformation:
\begin{equation*}
	\scalebox{1}{
		\begin{tikzpicture}
			\node (1a) at (0,0) {$\Hom{\CategoryC}{A}{B}$};
			\node (1b) at (3,0) {$FB$};
			\node (2a) at (0,2) {$\Hom{\CategoryC}{A}{A}$};
			\node (2b) at (3,2) {$FA$};
			
			\draw[->] (2a) to node[above]{$\eta_A$} (2b);
			\draw[->] (1a) to node[below]{$\eta_B$} (1b);
			\draw[->] (2a) to node[left]{$\Hom{\CategoryC}{A}{f}$} (1a);
			\draw[->] (2b) to node[right]{$Ff$} (1b);
	\end{tikzpicture}}
\end{equation*}
In particular, consider $A \xrightarrow{\Id{A}}A$. This is an element of 
$\Hom{\CategoryC}{A}{A}$, and $\Hom{\CategoryC}{A}{f}$ sends it to $f$ 
(which is by definition an element of $\Hom{\CategoryC}{A}{B}$) since it acts by 
post-composition and $f \circ \Id{A} = f$. Hence, if $\eta_A$ sends $\Id{A}$ to, say, 
$x \in FA$, then $\eta_B$ will have to send $f$ to $Ff(x)$, otherwise the square will not commute. 
This reasoning can be repeated for any $B$ and any $f$, it follows that the assignment 
$\eta_A (\Id{A}) = x$ \emph{completely determines} $\eta$. So we can have as many different 
$\eta$s as there are choices to which we can send $\Id{A}$. These are as many as the elements in $FA$, and so we have:
\begin{lemma}[Yoneda Lemma]
	There is a bijection\footnote{
		For readers versed in category theory: $\Hom{\CategoryC}{-}{-}$ is a contravariant 
		functor in the first component and a covariant functor in the second.
		It can be proven that the bijection is natural in $F$ and $A$.}
	\begin{equation*}
		\Nat{\Hom{\CategoryC}{A}{-}}{F} \simeq FA 
	\end{equation*}
	Where $\Nat{F}{G}$ denotes the set of natural transformations between any two functors $F,G$.
\end{lemma}
The consequences of Yoneda lemma are far reaching. In particular, the following corollary can be proven:
\begin{corollary}
	Given objects $A,B$ of $\CategoryC$, if it is\footnote{
		For readers versed in category theory: This bijection is 
		required to be natural in $F$. 
	} %
	\begin{equation*}
		\Nat{\Hom{\CategoryC}{A}{-}}{F} \simeq \Nat{\Hom{\CategoryC}{B}{-}}{F} 
	\end{equation*}
	for any functor $\CategoryC \xrightarrow{F} \Set$, then $A \simeq B$ 
	in $\CategoryC$.
\end{corollary}
As we already stressed, isomorphic objects in a category behave as if they were the same.
Repetitively applying Yoneda lemma one gets that
\begin{equation*}
	FA \simeq
	\Nat{\Hom{\CategoryC}{A}{-}}{F} \simeq \Nat{\Hom{\CategoryC}{B}{-}}{F} \simeq
	FB
\end{equation*}
And hence $A \simeq B$ if and only if $FA \simeq FB$ for any $\CategoryC \xrightarrow{F} \Set$.
So, with respect to whatever it is that we want to capture by defining a category $\CategoryC$, 
the Yoneda lemma affirms that we can \emph{completely} characterize an object $A$ by 
studying its images $FA$ for any functor $F$ to $\Set$.

We heavily rely on this 
to model attacker learning and interpreting functors 
to $\Set$ as testing procedures (Section~\ref{sec: security modelling}). Under this view, we summarize
Yoneda lemma as:
\begin{center}
	\emph{If two objects agree under any possible test we can perform,\\ then they behave the same.}
\end{center}
\printnomenclature
\end{document}